\documentclass[%
 aip,
 jmp,%
 amsmath,amssymb,
preprint,%
]{revtex4-1}
 
\usepackage{graphicx}
\usepackage{fullpage}
\usepackage{dcolumn}% Align table columns on decimal point
\usepackage{bm}% bold math

\begin{document}

\def\DoubleR{{\rm\bf R}}

\preprint{APS/123-QED}

\title{The stability of networks --- towards a structural dynamical systems theory}% Force line breaks with \\

\author{Michael Small}
%\affiliation{School of Mathematics and Statistics\\The University of Western Australia, Crawley, WA, Australia, 6009}
\email{michael.small@uwa.edu.au}
\author{Kevin Judd}
%\affiliation{School of Mathematics and Statistics\\The University of Western Australia, Crawley, WA, Australia, 6009}
\author{Thomas Stemler}
\affiliation{School of Mathematics and Statistics\\The University of Western Australia, Crawley, WA, Australia, 6009}

\date{\today}% It is always \today, today,

\begin{abstract}
The need to build a link between the structure of a complex network and the dynamical properties of the corresponding complex system (comprised of multiple low dimensional systems) has recently become apparent. Several attempts to tackle this problem have been made and all focus on either the controllability or synchronisability of the network --- usually analyzed by way of the master stability function, or the graph Laplacian.. We take a different approach. Using the basic tools from dynamical systems theory we show that the dynamical stability of a network can easily be defined in terms of the eigenvalues of an homologue of the network adjacency matrix. This allows us to compute the stability of a network (a quantity derived from the eigenspectrum of the adjacency matrix). Numerical experiments show that this quantity is very closely related too, and can even be predicted from, the standard structural network properties. Following from this we show that the stability of large network systems can be understood via an analytic study of the eigenvalues of their fixed points --- even for a very large number of fixed points. 
 \end{abstract}

\pacs{Valid PACS appear here}% PACS, the Physics and Astronomy
\maketitle

\section{Introduction}
\label{intro}

The principal aim of this paper is to ask what does the structure of a complex network tell us about the way that it will behave as a dynamical system. We are not the first to ask this question. Past attempts have focussed on either conditions to obtain a synchronous state, or controlling network dynamics to a target state. Our approach is to take the tools of dynamical systems theory and apply these to the network --- treating it as a very high dimensional dynamical system, with certain useful symmetric properties.  We present the results of this work in two parts. In Sec. \ref{computation} we present a computational study of the relationship between various network properties. Sec. \ref{examples} reverts to the basics of dynamical systems theory and presents, through several worked examples, direct application of these methods to network systems. In the remainder of this introduction and Sec. \ref{background} we briefly review the necessary background concepts.

Complex networks, and in particular, scale-free networks are widely described as ubiquitous throughout Nature. While this claim is reasonable, the widely utilized model for generating a scale-free network, preferential attachment \cite{aB99}, does introduce slight statistical bias for finite size realisations \cite{skinny}. For a finite network, constructed via preferential attachment, the connections from the last added nodes  are biased towards the hubs, and yet these nodes themselves have exceptionally low degree. This leads to finite scale free networks constructed with this method exhibiting negative assortativity. To overcome this we have proposed a form of {\em altruistic} attachment as an alternative\cite{pL11}: rather than connect directly to the hubs, new nodes are connected to random neighbors of the hubs. By doing this, the disassortatvity of preferential attachment is entirely mitigated \cite{pL11}.

Here, we choose to focus on the effect of this biased assortativity --- or disasortativity --- and ask how does the structural and dynamical behaviour of a strongly assortative (or disassortative) network differ from the archetypal preferential attachment model. To quantify the structural properties of a network we measure an ensemble of the usual suspects: node degree, diameter, assortativity, clustering and robustness (all to be briefly described latter). To measure dynamical behaviour we compute the leading eigenvalues of the fixed points of a network dynamical system (which is also described more precisely in the following sections). 

Against this battery of statistics we assess four different types of complex networks --- two standard and two of our own invention: (1) the standard preferential attachment model \cite{aB99}, (2) its small-world sibling \cite{dW98}, (3)  the altruistic attachment model briefly mentioned above, and (4) a ``skinny'' scale-free network designed to be not-small-world \cite{skinny}. From here on we refer to these four types of complex networks as: preferential attachment; small-world; altruistic attachment; and, skinny. For each of these four classes of networks we impose a link-exchange mechanism (which will also be described below) to incrementally alter the assortativity.  In the next section we introduce the necessary machinery.

\section{Networks and numerics}
\label{background}

In the following three subsection we introduce the necessary numerical techniques: statistical quantification of network structure (Sec. \ref{structure}), numerical measures of dynamical stability (Sec. \ref{dynamics}), and methods of link manipulation to modify network assortativity (Sec. \ref{links}). In all of what follows we consider unweighted undirected graphs of $N$ nodes represented by an $N\times N$ symmetric adjacency matrix $A$ such that $A_{ij}=1$ indicates the presence of a link between node-$i$ and node-$j$  ($A_{ij}=0$ otherwise). The main diagonal is zero ($A_{ii}=0$).

\subsection{Structure}
\label{structure}

The measures of network structure which we employ are fairly standard throughout the literature, we reiterate their description here briefly and refer the reader to \cite{mN10b} or the relevant sources for details.

\begin{itemize}
\item{node degree:} the average number of links for a node.
\item{diameter:} the median over all $i$ and $j$ of the shortest path between node-$i$ and node-$j$.
\item{assortativity:} the correlation coefficient between the degree of node-$i$ and the degree of the neighbors of node-$i$ --- computed over all $i$
\item{clustering:} the probability that two neighbors of node-$i$ are also neighbours --- i.e. the prevalence of triangles within the network.
\item{robustness:} the tendency for one of the properties of a network (usually diameter) to change with targeted or random deletion of nodes --- targeted deletion occurs when nodes which are deemed to contribute most to that property (high degree nodes,\footnote{Or, nodes with highest betweenness --- defined to be the fraction of shortest paths passing through that node. However, this is more costly to calculate.} for the case of diameter) are removed first.
\end{itemize}

\subsection{Dynamics}
\label{dynamics}

Let $\phi(x)$ define a dynamical system: $x'=\phi(x)$. In what follows we restrict our attention to the one dimensional case and consider a concrete example for the purposes of simulation: $\phi(x)=-x$ (we will expand on this example in Section \ref{1well}). Clearly, this system is globally stable with a single stable node\footnote{Here, {\em node} means a fixed point of a dynamical system with real eigenvalues --- nothing to do with the network.} at the origin $x_0=0$. Define the following network dynamical system:
\begin{eqnarray}
\label{sysA}
z' & = & \left(I+\epsilon A\right)\Phi(z)
\end{eqnarray}
where $I$ is the $N\times N$ identity and $\epsilon>0$ is a small positive coupling strength. The variable $z$ consists of $N$ sets of state variable of the system $x'=\phi(x)$ (since, here, $x\in\DoubleR$, $z\in\DoubleR^N$) and $\Phi(\cdot)=\phi(\cdot)\times\phi(\cdot)\times\ldots\times\phi(\cdot)$ is $N$ copies of the function $\phi(\cdot)$.

If $A$ is full rank, then the solutions of (\ref{sysA}) are $z_0\in\{z|\phi(z^{(i)})=0 \forall i\}$ where $z^{(i)}$ is the $i-th$ component of $z$. In the example we've chosen, this is trivial: $z=0$. The stability of this single $N$-dimensional fixed point is determined via the eigenvalues of the matrix $(I+\epsilon A)J_{\Phi(z_0)}$ where the Jacobian of $\Phi(z_0)$,  $J_{\Phi(z_0)}$,  is a diagonal matrix consisting of  the individual derivatives $\phi'(x_0)=-1$. For $\epsilon\ll 1$, $(1+\epsilon A)J_{\Phi(z_0)}\approx -I$ and these eigenvalues are all about $-1$. Hence, we are interested only in the largest eigenvalue of  $(I+\epsilon A)J_{\Phi(z_0)}$ (or equivalently, the smallest eigenvalue of  $(I+\epsilon A)$). For a given $A$ we can also determine the value of $\epsilon$ at which this first eigenvalue changes sign  --- at this point the system gains a single unstable direction, becomes a saddle, and (in the case of $\phi(x)=-x$) becomes unbounded.
 
\subsection{Links}
\label{links}

\begin{figure*}
\[\includegraphics[width=0.75\textwidth]{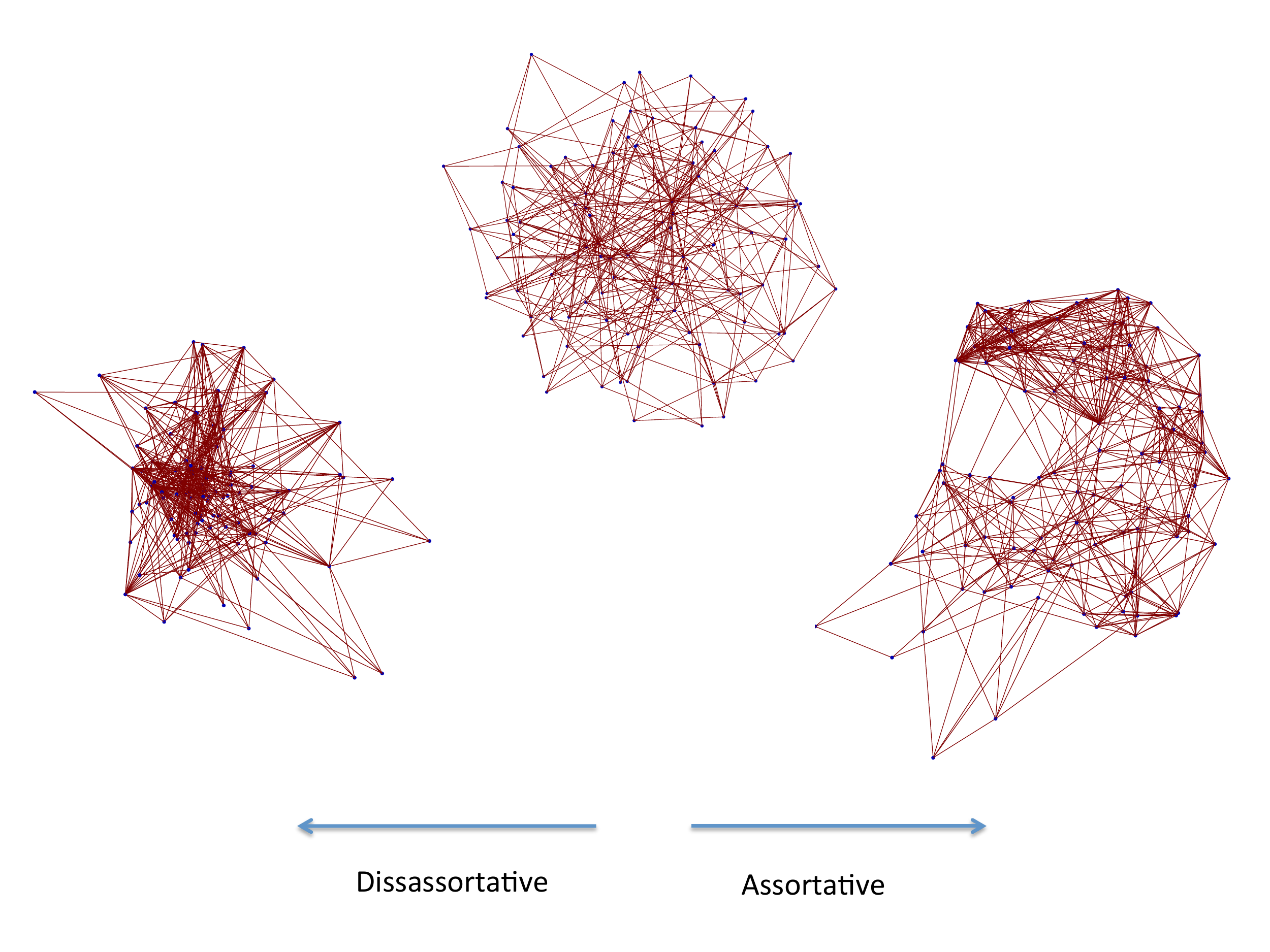}\]
\caption{{\bf Varying assortativity for preferential attachment networks:} The effect of increasing and decreasing assortativity via the link exchange mechanism described in the text. Starting with the small preferential attachment network ($100$ nodes) we increase and decrease assortativity with $50000$ applications of the link exchange mechanism to arrive at the assortative and dissassortative extrema shown below. At the dissassortative extreme, the network hubs are directly connected to the outliers. At the assortative end of the spectrum the nodes are strongly ordered according to their degree. The picture is very similar for the altruistic attachment model. The most notable feature of small-world and skinny scale free networks is the collapse of the network size with link-exchange. We choose very small networks as visualization of larger structures is less informative. }\label{zero}
\end{figure*} 

\begin{figure*}
\[\includegraphics[width=0.75\textwidth]{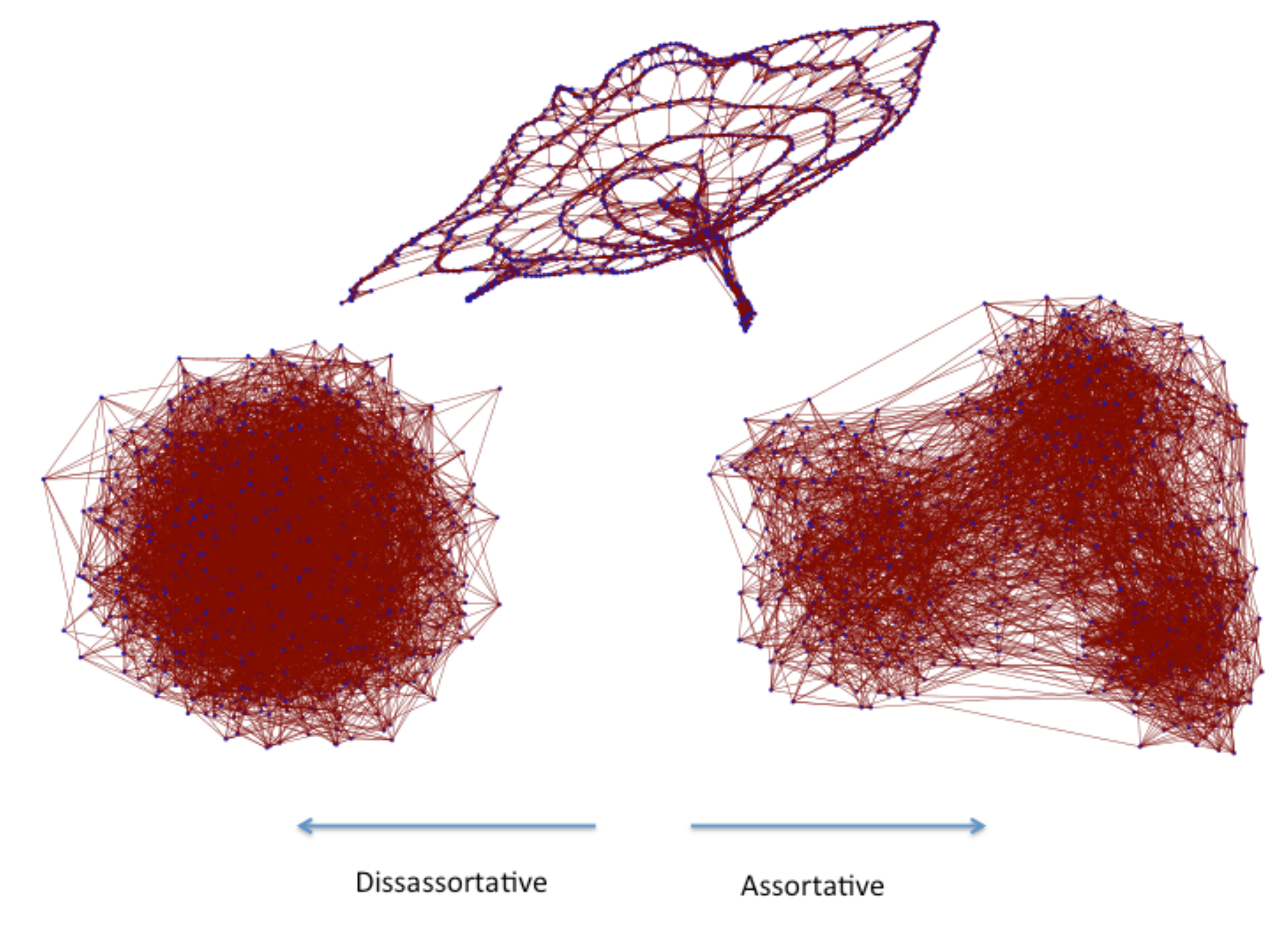}\]
\caption{{\bf Varying assortativity for the ``skinny'' scale free network:} The skinny scale free network is a rather pathological case --- it is a scale free network, but it is not small world. The top image depicts the global structure of such a network (of $1000$ nodes).  Starting from this network we increase and decrease assortativity with $50000$ applications of the link exchange mechanism to arrive at the assortative and dissassortative extrema shown in the lower portions of the figure.. At the dissassortative extreme, the network hubs are directly connected to the outliers and the network is highly compact. Multiple hubs are also loosely interconnected and are obscured in this projection. Conversely, at the assortative end we see community formation based on approximate node degree. Note that the network diameter is greatly decreased in either case.}\label{zero2}
\end{figure*} 

\begin{figure*}
\[\includegraphics[width=0.65\textwidth]{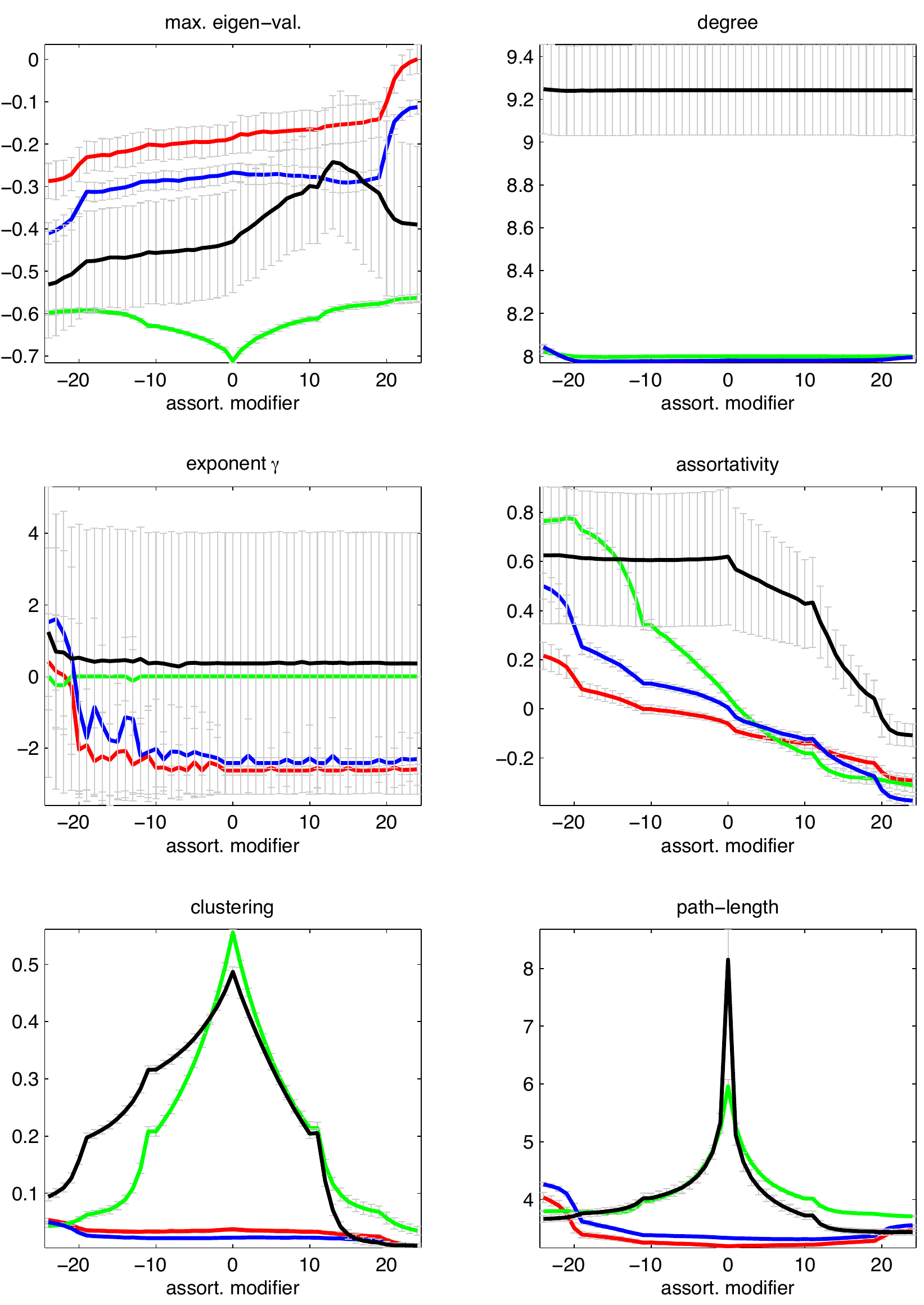}\]
\caption{{\bf Varying assortativity with a link-exchange mechanism:} The six panels depict computation of network properties described in the text after repeated application of link-exchange to either increase or decrease the assortativity. The four initial networks are: BA preferential attachment (red); the WS small-world model (green); ``altruistic attachment'' (blue); and, the ``skinny'' network (black). Mean and standard deviations from 52 trials are depicted. The horizontal axis (for every panel) is the extent to which the link-exchange mechanism was applied to increase (postive) or decrease (negative) the assortatitivity. Link-exchange was applied between 100 and 50000 times (the scale is not linear). }\label{first}
\end{figure*}

The four networks identified in the introduction exhibit a fairly wide range of assortativity --- two of them (the skinny network and the small-world network) are highly assortative. Nonetheless, how much more assortative (or disaasortative) can a network be? To explore a wider range of networks (ordered via assortativity) we implement the following simple link exchange mechanism:
\begin{enumerate}
\item pick two nodes $i$ and $j$ at random such that their degrees $d(i)$ and $d(j)$ differ
\item \label{switch}pick neighbors of node-$i$ and node-$j$, $\hat{i}$ and $\hat{j}$, such that the metrics
\[|d(\hat{i})-d(j)|\]
and 
\[|d(\hat{j})-d(i)|\]
are minimised
\item exchange links: disconnect node-$\hat{i}$ from node-$i$ and instead connect it to node-$j$, similarly disconnect node-$\hat{j}$ from node-$j$ and instead connect it to node-$i$
 \item repeat
 \end{enumerate}
 Note that the individual degrees of each node (and hence the degree distribution) are preserved, but since the degrees of the original nodes differ ($d(i)\neq d(j)))$ and link exchange mechanism attempts to produce strong links between nodes with similar degrees, the resultant network will tend to have a higher assortativity. 
 
Conversely, to decrease assortativity we do the reverse, step \ref{switch} is replaced with the following:
 \begin{enumerate}
\item[\ref{switch}'.] pick neighbors of these nodes $\hat{i}$ and $\hat{j}$ such that the metrics
\[|d(\hat{i})-d(i)|\]
and 
\[|d(\hat{j})-d(j)|\]
are minimised.
\end{enumerate}
 That is, instead of seeking to achieve a good match, we seek to destroy an existing good match.

Over time, these iterative schemes can be applied to significantly increase of decrease network assortativity. In the next section we present the results of our computational study of these networks. Figure \ref{zero} depicts the effect of these exchange mechanisms for the four types of networks considered herein.

\section{Assortativity and the dynamics of the network}
\label{computation}

\begin{figure*}
\[\includegraphics[width=0.85\textwidth]{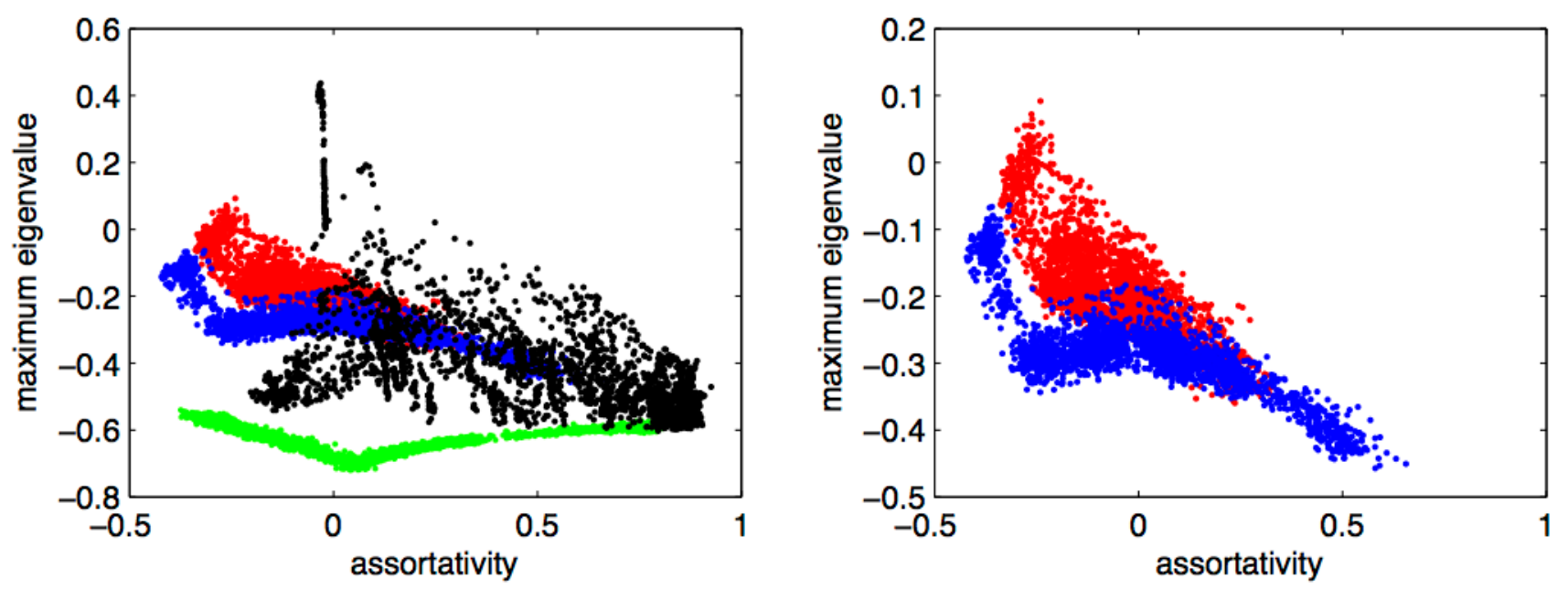}\]
\caption{{\bf Dynamic stability of the network as a function of network assortativity} Each network (with the same colour coding as Fig. \ref{first} is parameteerised by its assortativity and maximum eigenvalue and plotted here. The right hand panel re-plots results only for the preferential and altruistic attachment models. }\label{second}
\end{figure*} 

We have performed an extensive set of simulations to fully explore the parameter space delineated in the previous section. Figure \ref{first} highlights some of these results. For each of the four networks described in Sec. \ref{intro} we repeatedly apply the link-exchange mechanisms of Sec. \ref{links} to increase and decrease the network assortativity. For each network, we then compute the range of statistics described in Sec. \ref{structure} and \ref{dynamics}.

From Fig. \ref{first} we can draw several results. The link-exchange mechanism does have the desired effect on assortativity and has little effect on scale-free exponent (for networks from which this could be reliably estimated). Network degree is invariant --- as required by the link-exchange mechanism. The link modification seems to have a symmetric effect on both path-length and clustering of the small-world and skinny networks. Unlike the preferential and altruistic attachment models, both these networks are embeddable, or almost embeddable, in two dimensions (such that nodes that are close according to a Euclidean metric are also connected). The link modification scheme disrupts this property and hence decreases clustering and path-length --- as the networks become more random (exemplified in Fig. \ref{zero2}).  

\begin{figure*}
\[\includegraphics[width=0.85\textwidth]{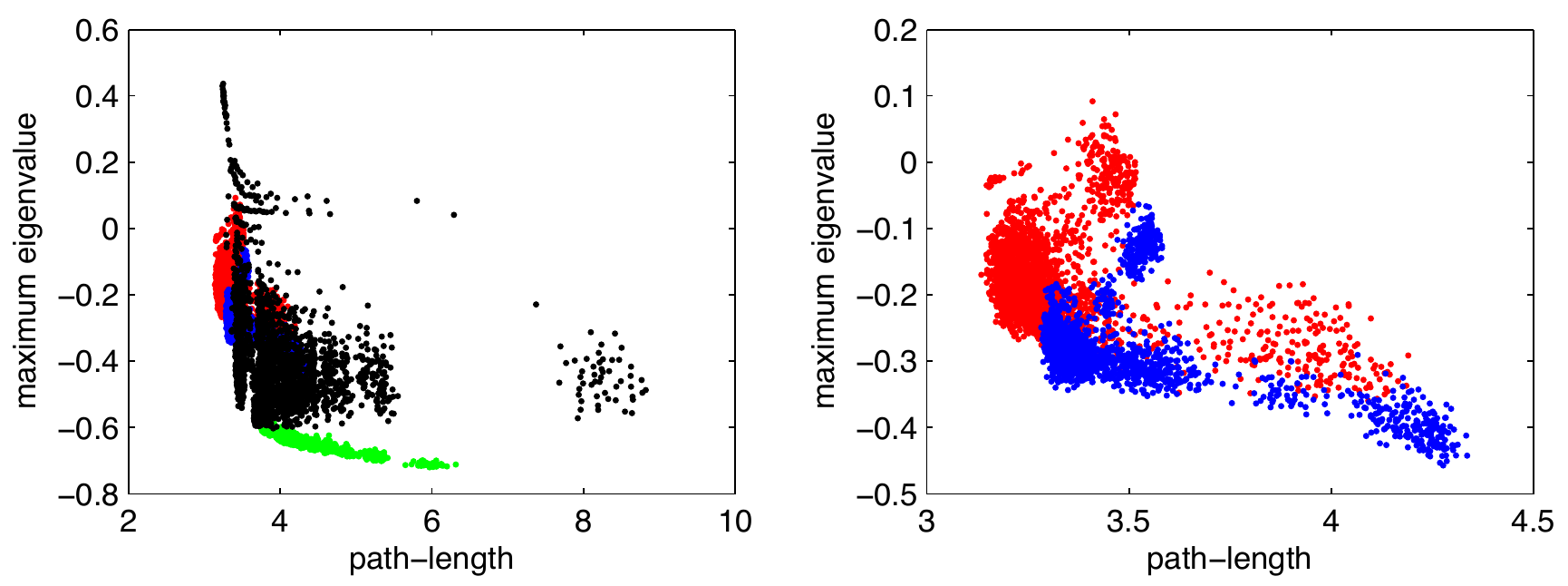}\]
\caption{{\bf Dynamic stability of the network as a function of network mean path-length (diameter)} Each network (with the same colour coding as Fig. \ref{first} is parameteerised by its mean path-length and maximum eigenvalue and plotted here. The right hand panel re-plots results only for the preferential and altruistic attachment models.}\label{third}
\end{figure*}

Finally, we observe that decreasing assortativity causes a nonlinear increase in the maximum eigenvalue of the resultant network for the scale-free networks --- with some notable exceptions. More assortative networks are more stable: and the effect is most pronounce for the most extremely assortative of disaasortative networks. In Fig. \ref{second} and Fig. \ref{third} we probe this relationship further and ask what can one learn about the dynamical structure of the network from examining the structural properties of path-length and assortativity. Of course, with all else being equal a marginally more assortative network will typically have a smaller diameter (path-length) as high degree nodes will tend to be wired together, rather than being connected to different regions of the network (the calculation of Fig. \ref{first} illustrates this point and the examples in Figs. \ref{zero} and \ref{zero2} demonstrate the opposite effect for larger -- fairly extreme -- changes in assortativity).

\begin{figure*}
\[\includegraphics[width=0.75\textwidth]{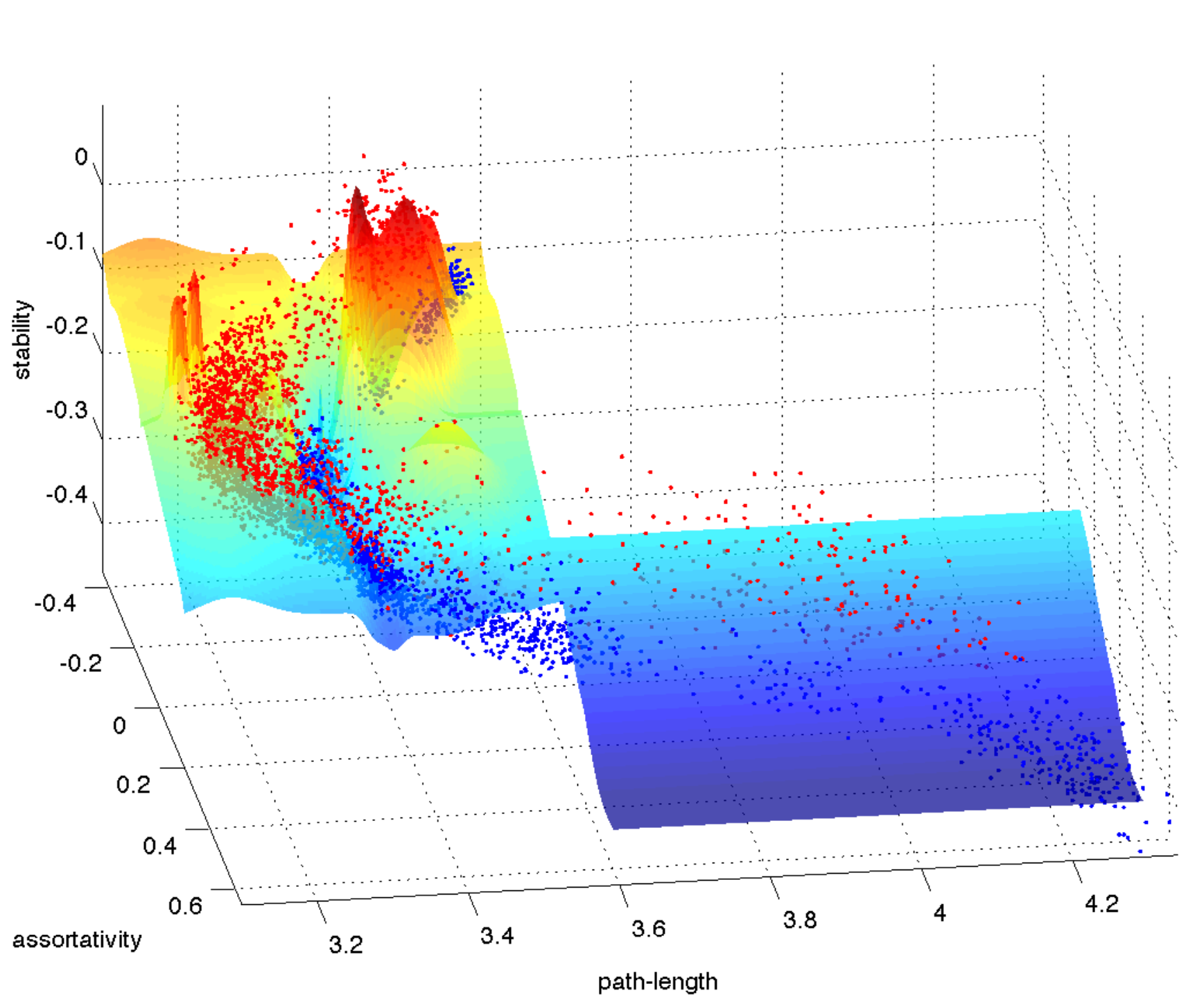}\]
\caption{{\bf Predicting stability} Contour plot of a nonlinear model prediction of stability as a function of assortativity and path-length overlaid on the data from Fig. \ref{second} and \ref{third}.}\label{fourth}
\end{figure*} 

In Fig. \ref{second} we see the effect of assortativity more clearly --- increasing assortativity decreases the maximum eigenvalue (making the network dynamically more stable). While widely distributed, one does observed that for a given value of assortativitiy the maximum eigenvalues are bounded above by a quantity which decreases linearly with assortativity.  At present we can offer no good explanation for this linear-in-assortativity upper bound on path length.  In Fig. \ref{third} we see a similar relationship between mean path length and maximum eigenvalue. Taking all three quantities together we find that the maximum eigenvalue can be predicted from a combination of assortativity and path-length --- in three dimensions the set of points depicted here approximates a smooth two-dimensional surface.  Figure \ref{fourth} depicts exactly this surface. We build a model (using the modeling procedures we have developed previously\cite{kJ95a,mdlnn} to predict stability (as defined above) from just assortativity and path-length. The error in these predictions (when applied to new data) has a mean of about $0.025$ ($\sim12\%$).

\begin{figure*}
\[\includegraphics[width=0.5\textwidth]{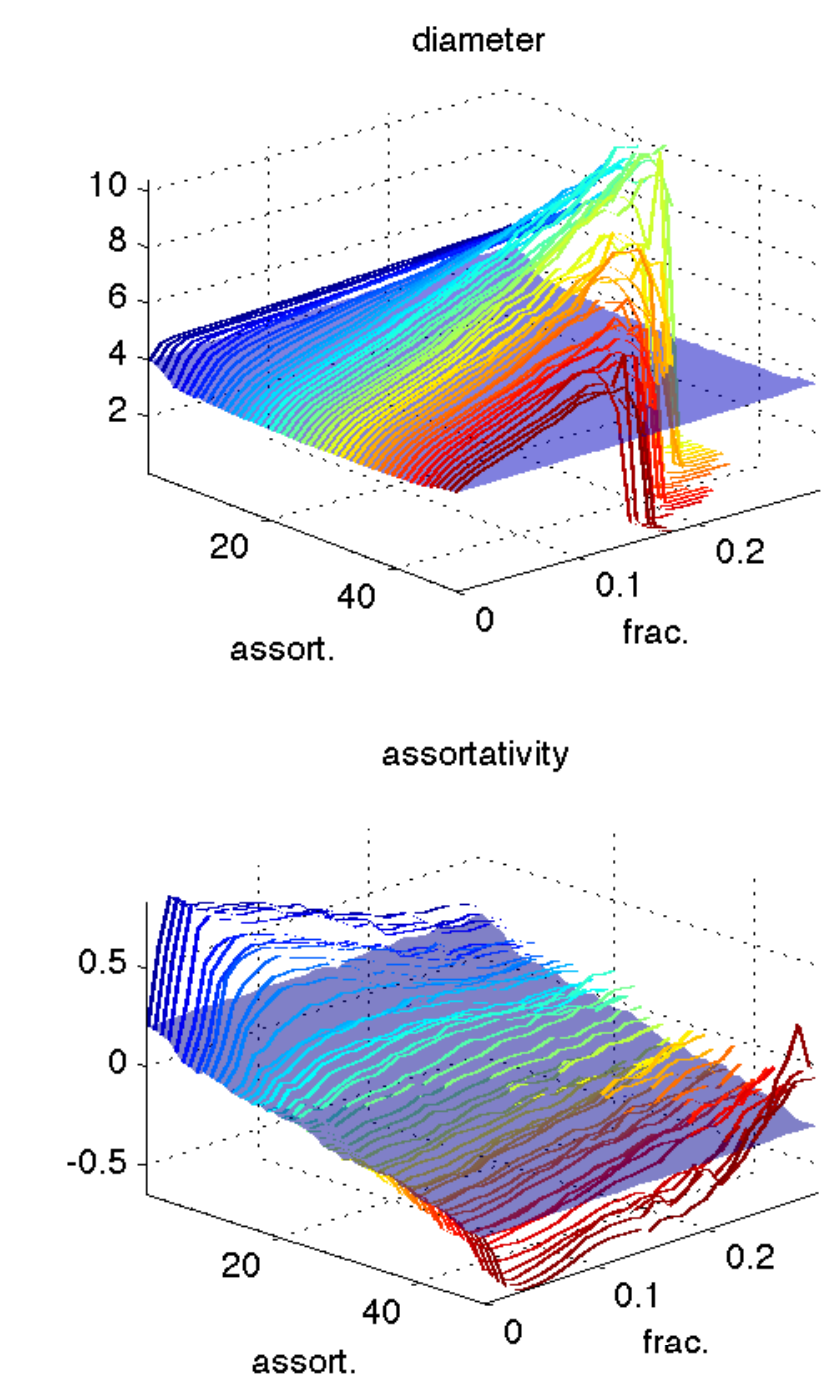}\]
\caption{{\bf Network robustness } Results for the preferential attachment model are shown - the altruistic attachment model produces qualitatively identical results, the other networks are more idiosyncratic. Again, we apply the link-rewiring scheme to change the assortativity of the network over a range of values --- the same range of values as depicted in Fig. \ref{first}: from highly assortative to disassortative (as the axis labelled ``assort.'' increases). For each such network we then perform targeted (the taper plot) and random (the translucent surface) node removal. Computed values of network diameter (mean minimum path-length) are shown on the upper panel and assortativity on the lower.}\label{fifth}
\end{figure*} 

Finally, in Fig. \ref{fifth} we compute the effect of changing the network assortativity on the robustness of the network. As shown in the upper panel, for all networks, targeted removal  results in a linear increase in network size (diameter) as the removal fraction increases --- up to the point where the network fractures. Random removal has very little effect on network size. This effect is robust across all levels of initial assortativity. Hence, the robust and fragile property of these networks is preserved. Conversely, targeted node removal has a very marked effect on network assortativity --- by removing a small fraction of nodes assortativity is very quickly attenuated.  Targetted link removal attenuates assortativity (making disassortative networks more disassortative and assortative networks more assortative) for all three classes of scale-free networks. Again, this is a curious observation and worthy of closer examination at a future time.  The effect is least pronounced in the skinny network and evident in the non-scale-free small world model only at the smallest fraction of link removals. With targeted removal of more links, small-world networks become less assortativity --- as the structure inherent in the network is disrupted. Across all types of networks, targeted link removal leads to an increase in clustering (since clustering indicates redundant links, and targeted link removal avoids such links).

The details of this computational study highlight several interesting connections among the various topological properties of complex networks --- and connections to network dynamics. The basic message from this work is that structure (and perhaps especially assortativity) matters.  In Sec. \ref{examples} we turn to looking more closely at exactly what this means for the dynamics of these systems.

\section{Analysis of Stability of Networks}
\label{examples}

To define the measure of  stability we used throughout the previous section we introduced a specific network dynamical system in Sec. \ref{dynamics}. We now revisit and expand on that definition here. In the subsequent subsections we work through several instructive examples.

As before, $x'=\phi(x)$ denotes the dynamics on each node and $A$ be the network adjacency matrix (hence $A_{ij}=1$ iff node-$i$ and node-$j$ are linked and $A_{ij}=0$ otherwise). Previously, we assumed that $x\in\DoubleR$, we now relax that assumption and allow the dynamics on the node to be a system of $k$ differential equations\footnote{Reformulation of this problem for the case of discrete dynamics and difference equations are left as exercises.}  $x\in\DoubleR^k$. Construct $\Phi(\cdot)=\phi(\cdot)\times\phi(\cdot)\times\ldots\times\phi(\cdot)$ from $N$ copies of $\phi$. 

For the sake of concreteness we assume that the nodal coupling is positive and bidirectional. Moreover, we assume that coupling occurs only amoung the first components of $x$ --- the only reason for this assumption is to simplify the construction of the $nk$-by-$nk$ coupling matrix $B$, below. Although the matrix $B$ is induced from $A$ under the restrictions on coupling just described, the analysis that follows will work equally well for an arbitrary matrix $B$. Let 
\begin{eqnarray}
\label{Bmatrix}
B_{ij}&=&\left\{\begin{array}{cc}
A_{(\frac{i}{k})(\frac{j}{k})} & \mbox{ if both $i$ and $j$ are integer multiples of $k$}\\
0 & {\rm otherwise}
\end{array}\right..
\end{eqnarray}
For one dimensional nodal dynamics, $B=A$, otherwise it is a sparse (sparser than $A$) matrix indicating coupling between the first component of the individual sub-systems, $(k-1)$ of $k$ rows and columns are all zero.

The global complex system dynamics can now be described by
\begin{eqnarray}
\label{syseq}
z' &=& (I+\epsilon B)\Phi(z)
\end{eqnarray}
where (again) $\epsilon>0$ is the coupling strength and the dynamics on the $n$-th node are described by $(z^{(kn)},z^{(kn+1)},\ldots,z^{((k+1)n-1})$.

The natural first step towards understanding a dynamical system such as (\ref{syseq}) is to find the fixed points, linearise them and then use this to characterize their stability. If $z_0$ is a fixed point we have $(I+\epsilon B)\Phi(z_0)=0$, i.e. $\Phi(z_0)\in\{u|{(I+\epsilon B)}u=0\}$ the nullspace of $(I+\epsilon B)$. If $(I+\epsilon B)$ is rank deficient, this then leads to a vector sub-space of $(I+\epsilon B)$ for which any solution of $\Phi(z)$ on that subspace is a fixed point. Such a non-trivial nullspace could (somehow) be factored out, so let us assume that the only solution is the trivial one and that $(I+\epsilon B)\Phi(z_0)=0$ implies $\Phi(z_0)=0$. Indeed, this is not a particularly onerous restriction, since a non-trivial nullspace implies nodes in $A$ (and $B$) that can be expressed as a linear combination of one another and could be trivially condensed to a single node. 

Hence, without loss of generality, the fixed points of (\ref{syseq}) are the fixed points of $x'=\phi(x)$. That is, $z_0$ is a fixed point iff  $\phi(z_0^{(kn)},z_0^{(kn+1)},\ldots,z_0^{((k+1)n-1})=0$ for $n=1,\ldots N$. We will come back to this later, but notice that this introduces quite a bit of symmetry --- if $\phi(x)=0$ has $m$ solutions then their are $m^N$ fixed points in (\ref{syseq}) situated at points in an  $N$-dimensional hyper-lattice.  Moreover, the location of the fixed points is independent of $B$ (and $A$) --- provided that no nodes in the network are redundant (i.e. $A$ has only the trivial nullspace).

Suppose that $\phi(x)=0$ has $m$ solutions: $a_1,a_2,\ldots a_m$ and call the set of these solutions ${\cal A}=\{a_1,a_2\ldots a_m\}$. Then $z_0$ is a fixed point of (\ref{syseq}) iff $z_0\in{\cal A}\times{\cal A}\times\ldots{\cal A}$. Hence, which fixed point we are interested can be identified with an $N$-tuple of integers from $1$ to $m$. Let $\ell(t)$ denote the value of the $t$-th component of that $N$-tuple --- hence the fixed point we are interested in from the  $t$-th subsystem is $a_{\ell(t)}$. Moreover, we can write $z_0=[a_{\ell(1)}\;a_{\ell(2)}\;\cdots\;a_{\ell(N)}]$.  Let $J_{\phi(a_i)}=b_i$ and then the stability of the fixed point $z_0$ is characterized by the eigenvalues and eigenvectors of the matrix 
\begin{eqnarray}
\label{eigeq}
(I+\epsilon B)\left[
\begin{array}{cccc}
b_{\ell(1)} & 0 & 0 & 0\\
0 & b_{\ell(2)} & 0 & 0\\
0 &0 &\ddots & 0 \\
0 &0 &0 &b_{\ell(N)}
\end{array}
\right].
\end{eqnarray}
Computationally this is all perfectly do-able, and the symmetry we alluded to earlier actually makes it quite simple for many systems of interest.

\subsection{Single wells: 1 fixed point}
\label{1well}

\begin{figure*}
\[\includegraphics[width=0.6\textwidth]{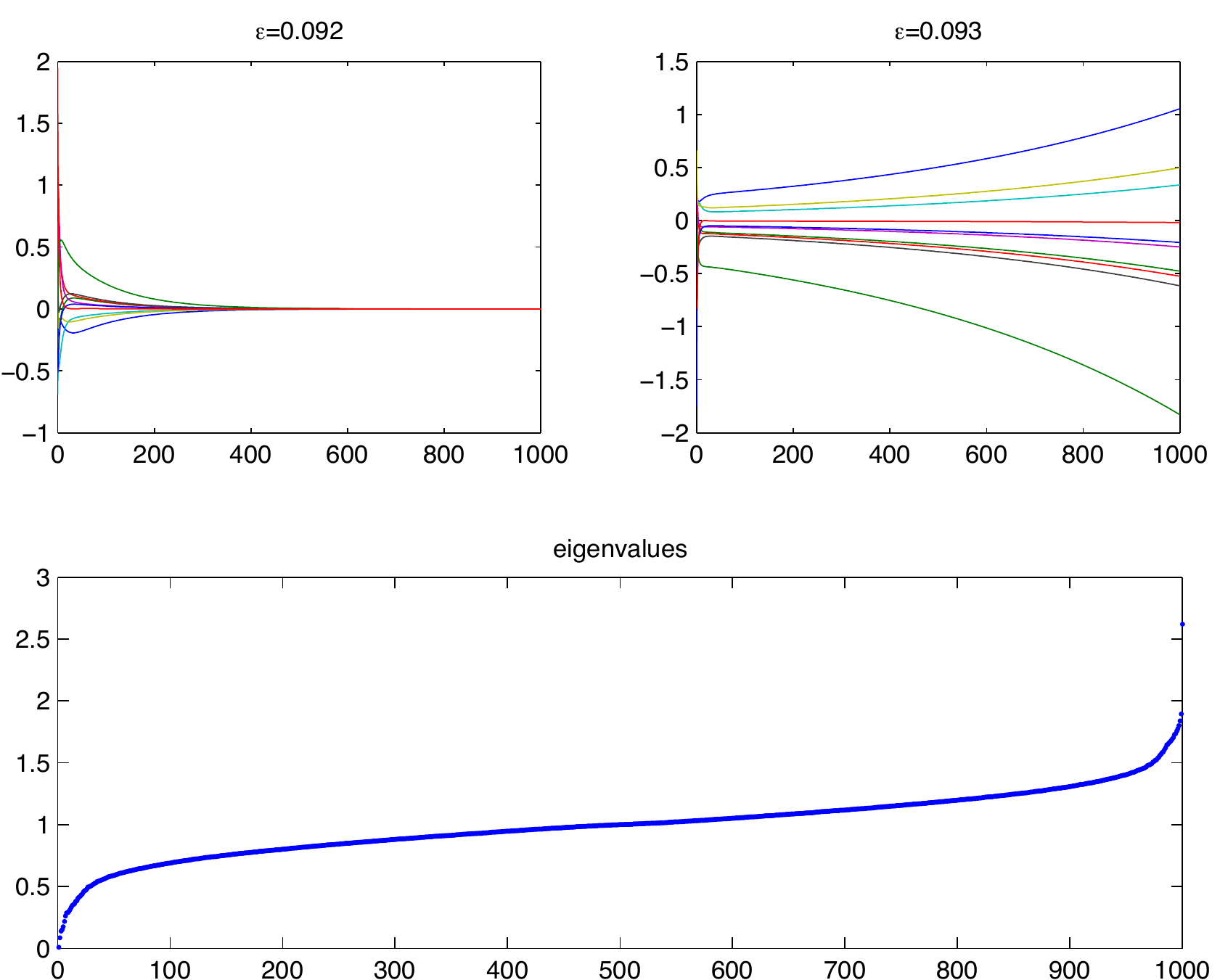}\]
\caption{{\bf Network robustness } Sample trajectories (the trajectories for the $10$ highest connected nodes are shown, all others are less divergent) for $A$ a preferential attachment scale-free network, $\phi(x)=-x$ and $N=1000$ nodes. Top-left is with $\epsilon=0.092$ and top-right is $\epsilon=0.093$. The eigenspectrum of $I+\epsilon A$ is shown in the lower plot (the eigenvalues of (\ref{syseq}) have the opposite sign because $\phi'(x)<0$). At $\epsilon=0.092$ all eigenvalues are negative, at $\epsilon=0.093$ one becomes positive. }\label{eg1}
\end{figure*} 

In Sec. \ref{dynamics} we considered the simplest possible non-trivial dynamical system. We do it again here. Let $\phi(x)=-x$. Hence, for $\epsilon=0$ the dynamics on each node consists of a one dimensional stable node at $x_0=0$. This one dimensional system means that $A=B$. As we increase $\epsilon\approx 0$, the matrix $(I+\epsilon A)>0$  has only positive eigenvalues. Hence the eigenvalues of (\ref{eigeq}) are all negative --- the system has a single stable node at the origin $z_0=0$. Eventually, for $\epsilon$ large enough the matrix (\ref{eigeq}) will gain first one and then increasingly more positive eigenvalues (two positive eigenvalues could emerge simultaneously given certain degenerate --- and symmetric --- arrangements of node in the network). But, as far as bounded dynamics are concerned, one unstable direction is sufficient: at this point the node becomes a saddle and all trajectories diverge. 

At first this may seem a little counter intuitive, and hence is interesting. Each node is very stable and yet by coupling these nodes together the system dynamics finds a way to climb out of this one dimensional potential well --- diverging to $\pm\infty$. Moreover, despite the universally positive coupling, some nodes diverge to $\infty$ while other go to $-\infty$., Figure \ref{eg1} depicts typical trajectories, confirming this behaviour and validating our choice to employ the largest eigenvalue of (\ref{eigeq}) as our measure of stability in Sec. \ref{computation}.

Finding the critical value $\epsilon_c$ such that the system's fixed point gains one unstable direction is also straightforward. Rewrite (\ref{eigeq}) in this case as $\left((-\epsilon B)-I\right)$ since $b_{\ell(i)}=\phi'(x_0)=-1$. The eigenvalues of this matrix are the solution of $\det{\left((-\epsilon B)-\left(1+\lambda\right)I\right)}=0$. Denote by $\lambda_B$ the eigenvalues of $B$ --- which satisfy $\det{(B-\lambda_BI)}$. Hence $\lambda=-1-\epsilon\lambda_B$ and the eigenvalues of our single potential well network system will all be negative if $1+\epsilon\lambda_B>0$. Hence the critical value $\epsilon_c$ for the node-saddle bifurcation is $\epsilon_c\lambda_{B}>-1$  for all the eigenvalues of $B$. Since $\epsilon>0$ this is always true of the positive eigenvalues. Hence we require  $\epsilon<-\frac{1}{\lambda_{B}}$. This is first violated by the negative eigenvalue of largest magnitude, i.e. $\lambda_{{B,\rm min}}$. Hence, the first bifurcation from node to saddle occurs at $\epsilon_c=-\frac{1}{\lambda_{B,{\rm min}}}$, where $\lambda_{B,{\rm min}}$ is the smallest eigenvalue of $B$. Subsequent transitions could be similarly computed, at least in principle. 
 
\subsection{Double wells: $3^n$ fixed points}
\label{2well}

\begin{figure*}
\[\includegraphics[width=0.6\textwidth]{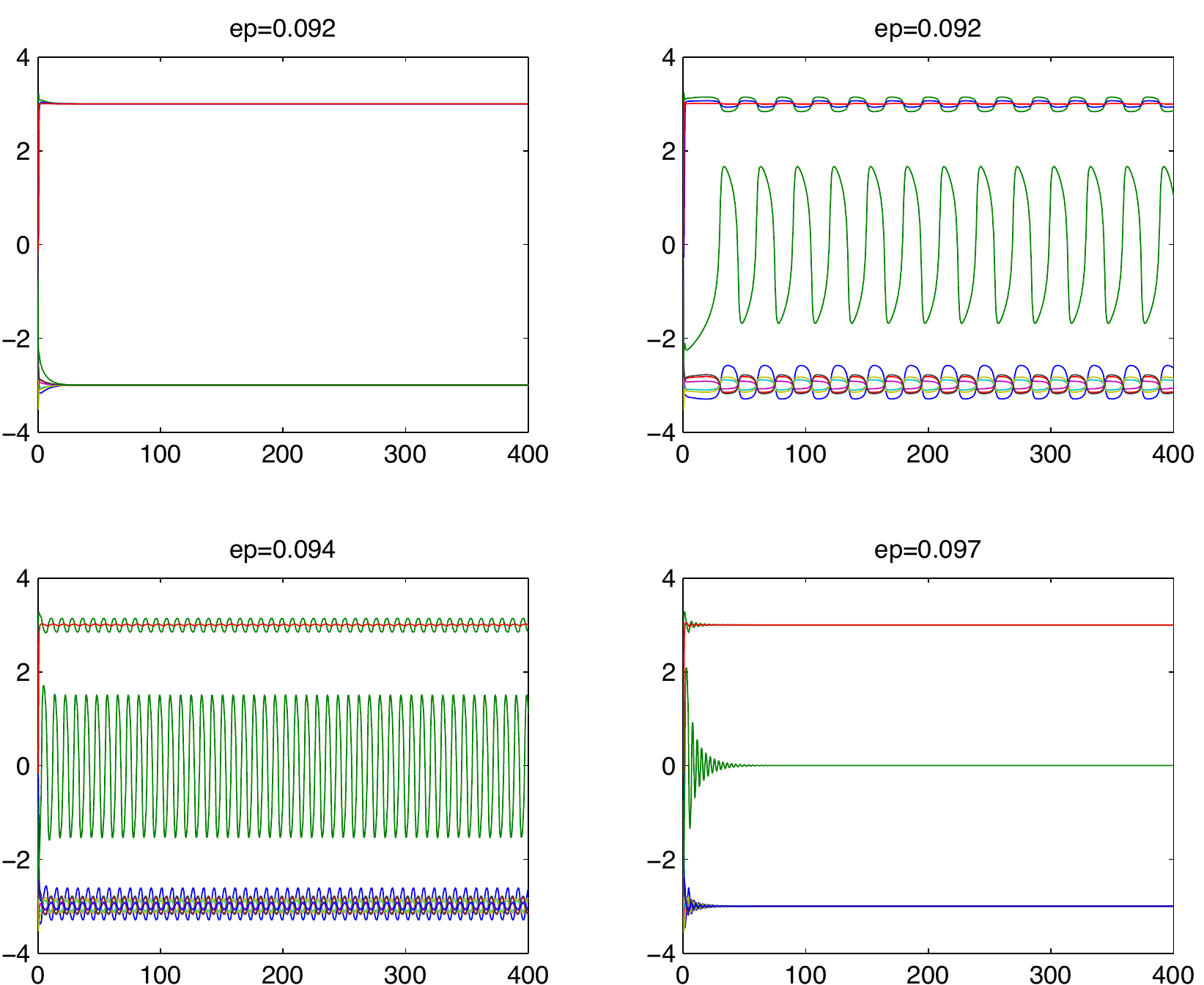}\]
\caption{{\bf Network robustness } Sample trajectories for $A$ a preferential attachment scale-free network (the same network as that used in Fig. \ref{eg1}), $\phi(x)=x(9-x^2)$ and $N=1000$ nodes. Top-left is with $\epsilon=0.092$; top-right is $\epsilon=0.093$; bottom left is $\epsilon=0.094$; and, bottom right is $\epsilon=0.097$. At $\epsilon=0.99$ the system becomes unbounded. The eigenspectrum is the same as that in Fig. \ref{eg1}.  }\label{eg2}
\end{figure*} 

The next situation is a little more interesting. For the time being we will stick to one dimensional dynamics, but we now introduce slightly more complex dynamics on the nodes. Let $\phi(x)=x(9-x^2)$. The system has three fixed points $x_0=\pm 3,0$. In the nomenclature introduced above $a_1=-3$, $a_2=0$, and $a_3=3$. The values of $\phi'$ are sufficient to characterize the stability of the fixed points $b_1=\phi'(a_1)=-18$, $b_2=\phi'(a_2)=9$, and $b_3=\phi'(a_3)=-18$: I.e. two stable and one unstable nodes.

Again, for a small $\epsilon>0$ the positive definiteness of $(I+\epsilon A)$ ensures that the fixed points behave independently of one another. Hence the fixed points of (\ref{syseq}) occur at $z_0$ where each component of $z_0$ is $-3$, $0$ or $3$. These are the vertices of a hypercube in $\DoubleR^N$ together with the mid-points of each edge and face and the centre (at the origin). There is one unstable node (corresponding to the fixed point at $0$) and $2^N$ stable nodes (each vertex of the cube). All other fixed points are saddles. As before, once $\epsilon$ increases the stable fixed points at the vertices of the cube gain unstable directions and become saddles --- computational simulation verifies that soon after this happens the system is no longer bounded. Because $\phi'(x_0)$ is a fixed constant at the stable nodes, this critical value is the same as in Sec. \ref{1well}:
 $\epsilon_c=-\frac{1}{\lambda_{B,{\rm max}}}$.

Figure \ref{eg2} confirms this analysis. Moreover, further computational probing shows, in Fig. \ref{eg2}, that the system develops first a stable periodic orbit around the origin and then a stable focus at $0$ --- that is, the fixed point which is initially unstable now becomes stable. Note that the fixed point depicted in Fig. \ref{eg2} is not the original in $\DoubleR^N$, but a point $z_0=[a_{\ell(1)}\;a_{\ell(2)}\;\cdots\;a_{\ell(N)}]$ such that there exists a unique $i$ such that $e_{\ell(i)}=0$ and for all other $j\neq i$, $e_{\ell(j)}=\pm 3$ --- that is, this is a point on one edge of the hypercube of fixed points described above.

 Finally, the more complex dynamics are a result of the correct orientation of stable and unstable manifold of saddle points and hence depend  sensitively on the choice of the matrix $A$ --- not just on the type of the network.

\subsection{Oscillators}

So, what happens if you start which oscillatory node dynamics? Can probably subsume this into the next example.

\subsection{Networked hyper-chaos: higher dimensional nodal dynamics}
\label{rosslers}

\begin{figure*}
\[\includegraphics[width=0.95\textwidth]{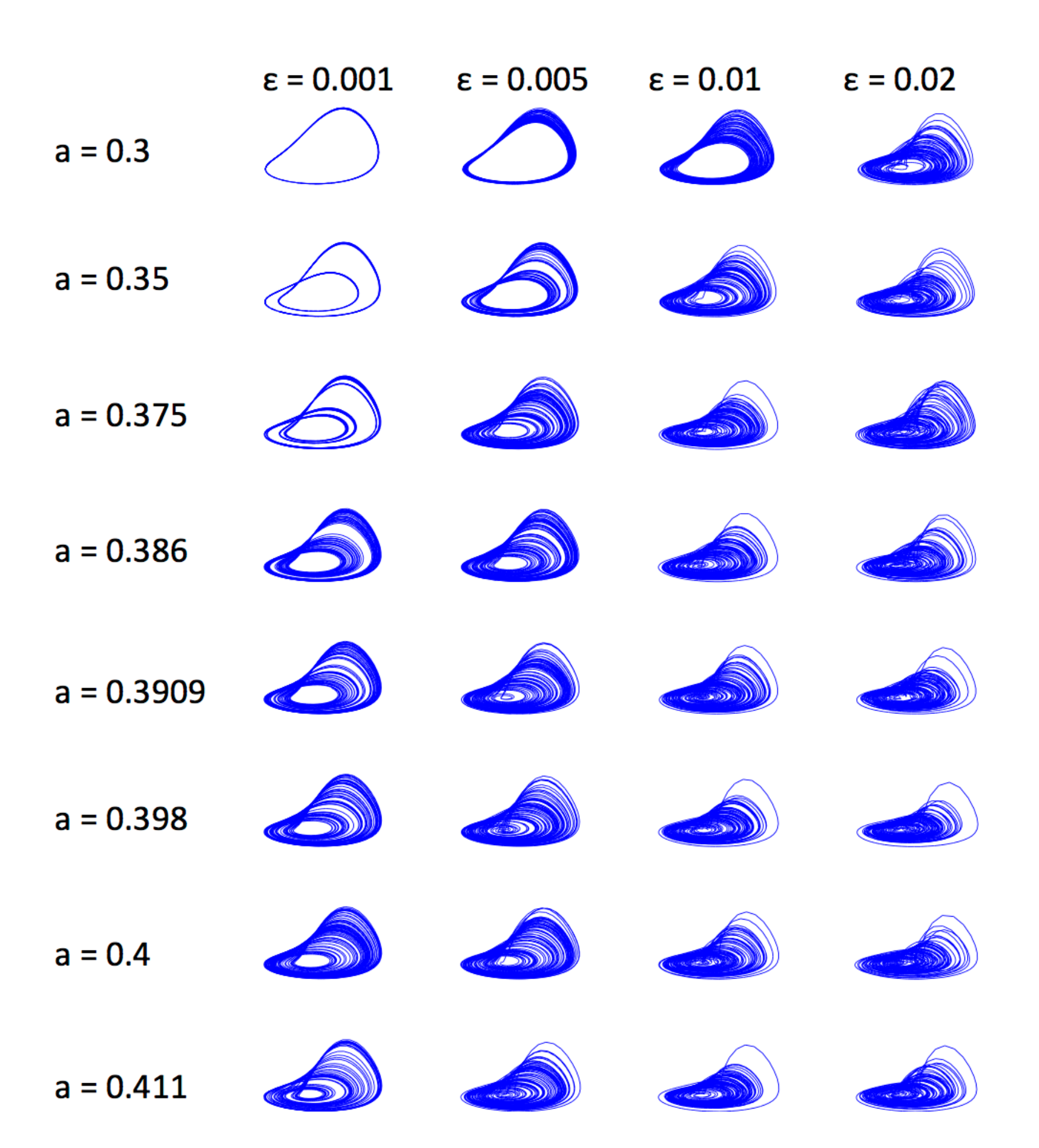}\]
\caption{{\bf Network of chaotic oscillators: } Sample trajectories for $A$ a preferential attachment scale-free network (the same network as that used in Fig. \ref{eg1}), $\phi(x)$ given by (\ref{rosseq}) and $N=1000$ nodes. Various values of the bifurcation parameter $a$ and the coupling strength $\epsilon$ are illustrated. Row-by-row, the uncoupled dynamics of the R\"ossler system at that value of $a$ exhibits: (1) limit cycle; (2) period 2; (3) period 4; (4) ``four-band'' chaos; (5) period 6; (6) ``broad-band'' chaos; (7) period 5; and (8) period 3. An expanded view of the system for  $a=0.398$ and $\epsilon=0.02$ is shown in Fig. \ref{rosseg2} }\label{rosseg1}
\end{figure*} 

\begin{figure*}
\[\includegraphics[width=0.95\textwidth]{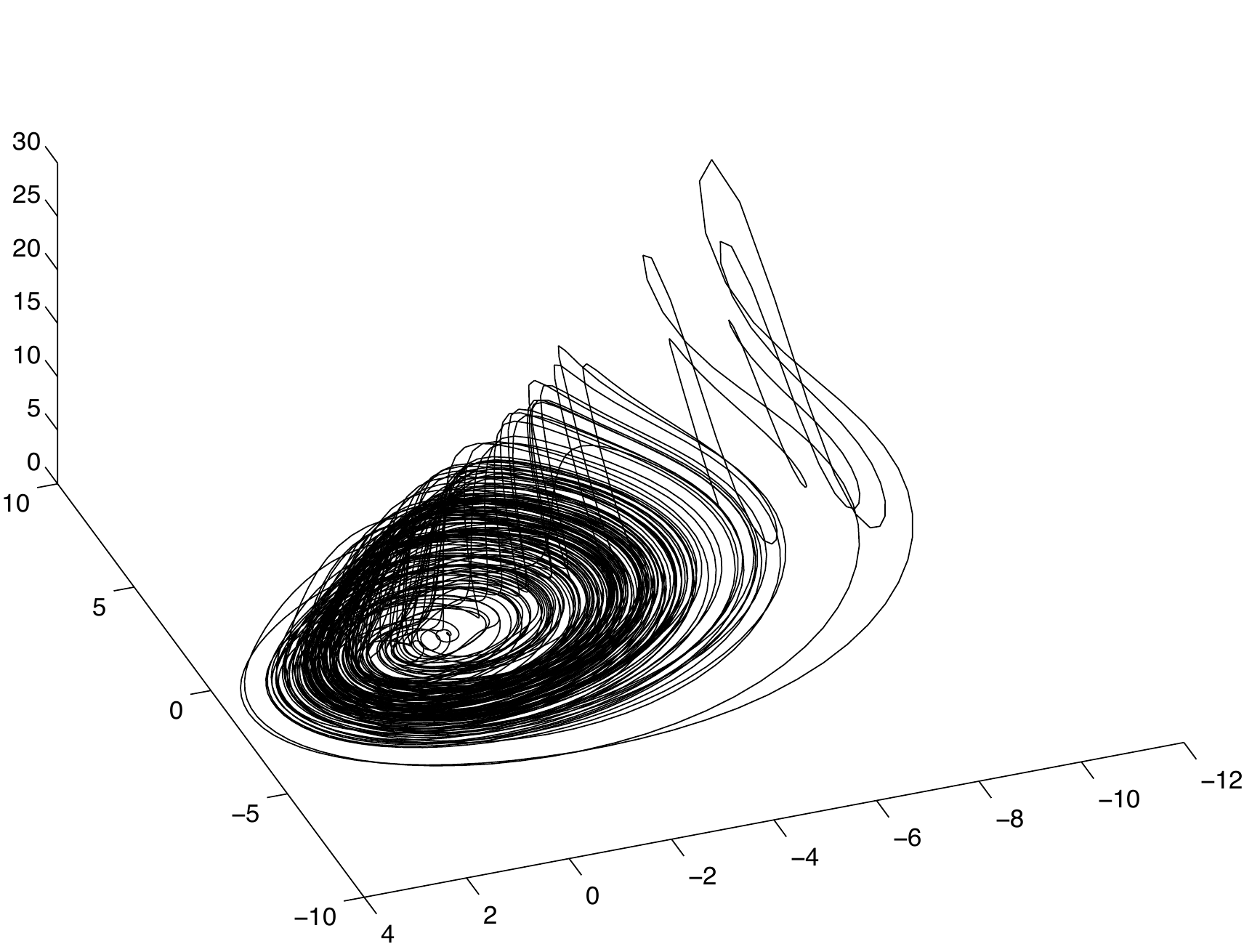}\]
\caption{{\bf Driven chaotic R\"ossler: } Sample trajectory for $A$ a preferential attachment scale-free network (the same network as that used in Fig. \ref{eg1}), $\phi(x)$ given by (\ref{rosseq}), $N=1000$ nodes, $a=0.398$ and $\epsilon=0.02$. We plot the dynamics on the largest hub for $10000$ time steps after a transient of $10000$ time step (sample time $0.1$). }\label{rosseg2}
\end{figure*} 

A lot of the notation we introduced at the beginning of  Sec. \ref{examples} is unnecessary if $\phi(x)$ is a scalar function. We now attempt to make better use of that notation. Let $x=(x_1,x_2,x_3)$ and 
\begin{eqnarray}
\label{rosseq}
\phi \left(\begin{array}{c}x_1 \\  x_2 \\ x_3\end{array}\right) 
& = & 
\left(
\begin{array}{c}
-x_2-x_3\\ x_1+ax_2\\ b+x_3(x_1-c)
\end{array}
\right) 
\end{eqnarray}
where $b=2$, $c=4$ and $a$ is a bifurcation parameter. This is, of course, the R\"ossler system where for $a=0.398$ the system exhibits broad band chaos and for smaller values of $a$ exhibits periodic dynamics and a period-doubling bifurcation via the archetypal stretching and folding mechanism. 

Although an analytical treatment of $N=1000$ coupled R\"ossler systems may be a little difficult, we can, at least, study the effect of such a coupling computationally and relate it to the dynamics we have already observed for simpler systems. Figures \ref{rosseg1} and \ref{rosseg2} summarise those calculations. Note that here $\epsilon\ll\epsilon_c$ --- the interesting dynamics highlighted here occur prior to any change in the stability of the fixed points.

\section{Conclusions}

It is not sufficient to collectivise all scale free networks and treat them equally. While some properties --- such as the so-called ``robust-yet-fragile'' character of these networks is preserved upon manipulation of assortativity. Other properties are sensitively dependent on the local connection between nodes. In particular,  more assortative networks are more dynamically stable. By merely changing the assortativity on a network one can progress from a regime of stable equilibrium dynamics to global instability. Nonetheless, we do find that, to a very large degree, stability of a network can be predicted from just network diameter and path-length --- two global properties of connectivity that are, in turn, strongly affected by assortativity.

Treating the networks as dynamical systems we are able to make use some basic results from dynamical systems theory to obtain some level of understanding of the evolution of complex dynamics in these systems. We see that even for a fairly large dimensional systems, if the dynamics on then individual nodes are identical then the system can be understood in terms of those fixed points. Moreover, the initial transition from a stable node to a dynamical saddle does not depend on $\phi'(x_0)$, but can be determined directly from only the smallest eigenvalue of $A$.

In computational simulations of large network systems we observe interesting effects of the compounding dynamics. Single well-potential functions can be forced to ``climb'' away from the stable equilibrium. Unstable equilibira in double wells are stabilised, and for chaotic systems the dynamical coupling appears to intensify the complexity (and possible the chaotic-ness) of the nodal dynamics. 

Finally, we computationally demonstrate a strong dependence of the system dynamics (characterized by this smallest eigenvalue of $A$) on the median path-length and assortative of the network. Moreover, we have uncovered an apparent linear-in-assrotativity upper bound on the dynamical stability of these network systems.

\bibliography{../../manuscripts//bibliography}
\bibliographystyle{abbrv}


\end{document}